\documentclass[bibyear]{aastex}
\usepackage{spr-astr-addons}
\usepackage{url}
\usepackage{graphicx}
\usepackage[fleqn]{amsmath}
\RequirePackage{color}

\begin{document}

\title{Effects of suprathermal electrons on the proton temperature anisotropy in space plasmas: Electromagnetic ion-cyclotron instability}

\slugcomment{Not to appear in Nonlearned J., 45.}
\shorttitle{Electromagnetic ion cyclotron instability}
\shortauthors{Shaaban et al.}

\author{S.~M.~Shaaban\altaffilmark{1,2}} \and \author{M.~Lazar \altaffilmark{3,4}} \and \author{S.~Poedts \altaffilmark{1}} \and \author{A.~Elhanbaly\altaffilmark{2}}

\altaffiltext{1}{Centre for Mathematical Plasma Astrophysics, Celestijnenlaan 200B, B-3001 Leuven, Belgium.} \email{shaaban.mohammed@student.kuleuven.be}
\altaffiltext{2}{Theoretical Physics Research Group, Physics Department, Faculty of Science, Mansoura University, 35516, Egypt.}
\altaffiltext{3}{Royal Belgian Institute for Space Aeronomy 3-Avenue Circulaire, B-1180 Brussels, Belgium.}
\altaffiltext{4}{School of Mathematics and Statistics, University of St Andrews, St Andrews, Fife, KY16 9SS, U.K.}

\begin{abstract}
In collision-poor plasmas from space, e.g., the solar wind and planetary magnetospheres, 
the kinetic anisotropy of the plasma particles is expected to be regulated by the kinetic 
instabilities. Driven by an excess of ion (proton) temperature perpendicular to the magnetic 
field ($T_\perp >T_\parallel$), the electromagnetic ion-cyclotron (EMIC) instability is fast 
enough to constrain the proton anisotropy, but the observations do not conform
to the instability thresholds predicted by the standard theory for bi-Maxwellian models of 
the plasma particles. This paper presents an extended investigation of the EMIC instability in the 
presence of suprathermal electrons which are ubiquitous in these environments.
 The analysis is based on the kinetic (Vlasov-Maxwell) theory assuming that both species, 
protons and electrons, may be anisotropic, and the EMIC unstable solutions are
derived numerically providing an accurate description for conditions typically encountered 
in space plasmas. The effects of suprathermal populations are triggered by the electron 
anisotropy and the temperature contrast between electrons and protons. For certain 
conditions the anisotropy thresholds exceed the limits of the proton anisotropy measured in
the solar wind considerably restraining the unstable regimes of the EMIC modes.
\end{abstract}


\section{INTRODUCTION}

Different physical mechanisms, which are at work in the solar wind and planetary magnetospheres, e.g., 
the solar wind expansion, compression of the magnetic field, may lead to large deviations 
from isotropy of the velocity distributions of  plasma particles. Indeed, the in-situ
measurements reveal the anisotropy of electrons and protons, which, for instance, exhibit different temperature 
components $T_{\perp } \ne T_{\parallel }$ relative to the local magnetic field $\boldsymbol{B_{0}}$ 
 \citep{Hellinger2006, stverak2008}. However, these deviations from isotropy are
not large, and in the absence of collisions only the resulting instabilities may scatter 
particles back towards a quasi-equilibrium state and prevent the anisotropy to grow 
indefinitely \citep{Gary1994, Gary2001, Bale2009}. Thus, in collisionless plasmas linear
dispersion theory predicts that protons (ions) with an excess of perpendicular temperature 
$T_{p,\perp }>T_{p,\parallel }$ may drive the electromagnetic ion cyclotron (EMIC) instability 
\citep{Gary1993}. According to \cite{Kennel1966}, this instability has a maximum growth 
rate at parallel propagation $\left(\boldsymbol{k}\times \boldsymbol{B_{0}}=0\right)$, where the
EMIC modes are left-hand (LH) circularly polarized.

In competition with mirror instability that develops for the same
conditions, the EMIC modes grow faster \citep{Gary1976}, but the
mirror thresholds shape better the limits of the proton anisotropy
observed in the solar wind at 1 AU \citep{Hellinger2006}. One possible
explanation for this disagreement may reside in the fact that the EMIC fluctuations also dissipates 
faster due to resonant cyclotron interactions with protons  \citep{Bale2009}. 
Otherwise, this disagreement may be a result of the limitations in the approach, 
which ignores either the interplay of protons with other species (e.g., alpha particles,
electrons) or the presence of suprathermal (Kappa-like) populations.
The presence of alpha particles changes the dispersive
properties of the plasma and introduces the alpha cyclotron
instability \citep{Hellinger2005}. For certain conditions, e.g.,
protons and alpha particles with the same temperatures and
temperature anisotropies, numerical simulations predict EMIC
thresholds with a better alignment to the limits of proton
anisotropy in the fast winds but not in the slow winds
\citep{Matteini2007}. On the other hand, the instability conditions are found to be markedly influenced
by the electrons with anisotropic temperatures, and their effect is enhanced
by the electron-proton temperature ratio. However, the instability thresholds obtained in this case
do not indicate better constrains for the proton anisotropy in the solar wind \citep{Shaaban2015}.

The electrons have an impact on proton instability by changing the
wave phase velocity, which may increase the number of resonant
protons and, implicitly, enhance the growth rate of the instability
\citep{Kennel1968}. Mutual effects of electrons and ions have
already been studied for conditions favorable to the firehose instability \citep{Kennel1968, 
Michno2014} or the EMIC instability \citep{Shaaban2015},
but the influence of suprathermal populations was neglected in these cases. 
In the present paper we demonstrate the existence of new regimes of this instability
triggered by the suprathermal electrons, which are ubiquitous in space
plasmas. Enhanced by these suprathermal populations, the high energy tails of the 
electron velocity distribution functions (VDFs) measured in the solar wind (at various heliographic coordinates) 
are well described by the Kappa distribution functions \citep{Vasyliunas1968, Feldman1975, 
Maksimovic1997}, which are nearly Maxwellian at low energies and decrease as power-laws at high 
energies. In the last decades, this realistic model has proved to be a veritable tool of modelling 
particle velocity distributions, replacing or complementing the standard Maxwellian distribution function when describing 
space pasma systems out of thermal equilibrium \citep{Hellberg2005, Hellberg2009, Pierrard2010,Livadiotis2013}.

The EMIC instability conditions are therefore expected to be markedly altered by the deviations from 
thermodynamic equilibrium shown by the electron distributions in the solar wind, which may cumulate
the effects of suprathermal populations and the anisotropic temperatures \citep{Maksimovic2005, 
stverak2008}. To include these effects in the analysis,  we assume the electrons bi-Kappa 
distributed \citep{Summers1991}, a model widely invoked to describe the kinetic instabilities 
in space plasmas \citep{Mace2011, Kourakis2012, Lazar2012, Henning2014, Lazar2015b}. In Section 
2 we introduce the distribution models for electrons and protons, and provide the dispersion 
relation for the EMIC modes. The unstable solutions as wave-number spectra of the growth 
rates and wave-frequencies are obtained numerically, enabling also to derive the anisotropy 
thresholds close to the marginal stability. In Section 3 we focus on the EMIC instability in 
the conditions relevant for the solar wind and planetary magnetospheres. The conclusions of 
the present  study are presented in the last section.

\section{GOVERNING DISPERSION RELATION}

First, we introduce the models for the VDFs of the principal
plasma components, the electrons (subscript $e$) and the protons (subscript
$p$). In the unperturbed state the anisotropic protons are assumed
bi-Maxwellian
\begin{equation}
F_{p}\left( v_{\parallel },v_{\perp }\right) =\frac{1}{\pi
^{3/2}u_{p ,\perp }^{2} u_{p ,\parallel }}\exp \left(
-\frac{v_{\parallel }^{2}} {u_{p,\parallel }^{2}}-\frac{v_{\perp
}^{2}}{u_{p ,\perp }^{2}}\right),   \label{e1}
\end{equation}
where thermal velocities $u_{p,\parallel, \perp}$ are defined by the 
components of the anisotropic temperature
\begin{equation}
T_{p,\parallel}^M=\frac{m}{k_B}\int d\textbf{v} v_\parallel^2
F_p(v_\parallel, v_\perp)=\frac{m u_{p ,\parallel}^2}{2 k_B}
\end{equation}
\begin{equation}
T_{p,\perp}^M=\frac{m}{2 k_B}\int d\textbf{v} v_\perp^2
F_p(v_\parallel, v_\perp)=\frac{m u_{p ,\perp}^2}{2 k_B}.
\end{equation}

Enhanced by the suprathermal populations, the anisotropic electron
distributions are described by a bi-Kappa VDF \citep{Summers1991}

\begin{equation}
     \begin{aligned}
F_e= \frac{1}{\pi ^{3/2}\theta _{e, \perp }^{2}\theta _{e, \parallel}}&\frac{\Gamma\left( \kappa_e +1\right) }{\Gamma \left( \kappa_e -1/2\right) }\\
       &\times\left[ 1+\frac{v_{\parallel }^{2}}{\kappa_e \theta _{e, \parallel }^{2}}+\frac{v_{\perp }^{2}}{\kappa_e \theta _{e, \perp }^{2}}\right] ^{-\kappa_e -1},\label{2}
     \end{aligned}
\end{equation}
which is normalized to unity $\int d^{3}vF_e=~1$, and is written in terms of 
thermal velocities $\theta_{e,\parallel, \perp}$ defined by the components of the effective 
temperature (for a power-index $\kappa_e >3/2$)
\begin{equation}
T_{e,\parallel}^K=\frac{2 \kappa_e}{2 \kappa_e-3}\frac{m_e }{2 k_B}\theta_{e ,\parallel}^2
\end{equation}
\begin{equation}
T_{e,\perp}^K=\frac{2 \kappa_e}{2 \kappa_e-3}\frac{m_e }{2
k_B}\theta_{e ,\perp}^2. \label{e6}
\end{equation}

Without suprathermals, the electrons become bi-Maxwellian distributed ($\kappa \to  \infty$) and the
components of their temperature reduce to
\begin{equation}
\lim_{\kappa\to\infty} T_{e,\parallel, \perp}^K= \frac{m_e }{2
k_B}\theta_{e ,\perp}^2 = T_{e,\parallel,\perp}^M.
\end{equation}
In the presence of suprathermal populations quantified by a finite power-index $\kappa$, the electron
temperature is enhanced  \citep{Lazar2015b}
\begin{equation}
T_{e,\parallel, \perp}^K=\frac{2 \kappa_e}{2 \kappa_e-3}T_{e,\parallel,\perp}^M
> T_{e,\parallel,\perp}^M,
\end{equation}
and so is the electron plasma beta ($\beta_e = 8 \pi n_e k_B T_e
/B_0^2$)
\begin{equation}
\beta_{e,\parallel, \perp}^K=\frac{2 \kappa_e}{2 \kappa_e-3}
\beta_{e,\parallel,\perp}^M > \beta_{e,\parallel,\perp}^M.
\label{e8}
\end{equation}
leading to the conclusion that dispersion properties and, implicitly, the instability conditions 
must also change. We have invetigated the effects of suprathermal populations on the plasma waves and instabilities using two alternative approaches, which assume
the effective temperature to be either constant \citep{Lazar2011, Mace2011, Lazar2012}, 
or increasing with the increase of suprathermal populations, i.e., with decreasing 
the power-index $\kappa$ \citep{Leubner2000, Lazar2015b, Lazar2016}.
Being more convenient computationally the approach with a $\kappa$-independent temperature  
has been widely invoked in similar theoretical predictions of plasma instabilities. 
However, \cite{Lazar2015b} have recently shown and \cite{Lazar2016} have extended the comparative 
analysis to add supplementary arguments that a Kappa model with the effective temperature
increasing with growing suprathermal population, i.e., with decreasing value of the power index $\kappa$,
reproduces much better the Maxwellian core in the limit  $\kappa \to \infty$, and thus enables a more
realistic characterization of the suprathermal populations and their destabilizing effects.

For a collisionless and homogenous electron-proton plasma
described by the distributions (\ref{e1})--(\ref{e6}), the
dispersion relations for the electromagnetic modes propagating
parallel to the stationary magnetic field read
\begin{equation}
     \begin{aligned}
D^{\pm }\left( k,\omega \right)=1-\frac{c^{2}k^{2}}{\omega ^{2}}+\frac{\omega _{p,p }^{2}}{\omega ^{2}}\left[\frac{\omega}{k u_{p,\parallel}} Z\left( \xi _{p}^{\pm }\right)\right.
&\\\left.+ \left( A_{p}-1\right)\left\{ 1+\xi _{p }^{\pm}Z\left(\xi _{p}^{\pm }\right) \right\} \right]+\frac{\omega _{p,e }^{2}}{\omega ^{2}}\\
\times \left[\frac{\omega }{k\theta _{\parallel ,e }}Z_\kappa\left(g_{e}^{\pm }\right)+\left( A_{e}-1\right)\right.\\
\times \left.\left\{ 1+\frac{\omega \pm \Omega _{e}}{k\theta _{e,
\parallel}}Z_\kappa\left( g_{e}^{\pm }\right) \right\}\right]=0,
\label{e9}
     \end{aligned}
\end{equation}
where $\omega $ is the wave-frequency, $k$ \ is the wave-number, $c$
is the speed of light in vacuum, $\Omega _{\alpha }=q_{\alpha
}\mathbf{B}_{0}/m_{\alpha }c$ is the gyrofrequency
(non-relativistic), $\omega _{p,\alpha }^{2}=4\pi n_{\alpha
}e^{2}/m_{\alpha }$ is the plasma frequency of different species
$(\alpha=p,e)$, $\pm $ denote, respectively, the circular
right-handed (RH) and left-handed (LH) polarization, and $A_{\alpha
}= T_{\alpha, \perp}/T_{\alpha, \parallel}$ is the temperature
anisotropy. In the direction parallel to the magnetic field, the EMIC modes
decouples from the electrostatic oscillations, and their instability exhibits maximum growth
rates \citep{Kennel1966}. The dispersion relation is obtained in
terms of the plasma dispersion function \citep{Fried1961}
\begin{equation}
Z\left( \xi _{p}^{\pm }\right) =\frac{1}{\pi ^{1/2}}\int_{-\infty
}^{\infty }\frac{\exp \left( -x^{2}\right) }{x-\xi _{p}^{\pm }}dt,\
\ \Im \left( \xi _{p}^{\pm }\right) >0,  \label{6}
\end{equation}
of argument
\begin{equation*}
\xi _{p}^{\pm }=\frac{\omega \pm \Omega _{p}}{ku_{p,\parallel,}},
\end{equation*}
and the modified (Kappa) dispersion function \citep{Lazar2008}

\begin{equation}
     \begin{aligned}
     Z_\kappa\left( g_{e}^{\pm }\right) =&\frac{1}{\pi ^{1/2}\kappa_{e}^{1/2}}\frac{\Gamma \left( \kappa_{e} \right) }{\Gamma \left(\kappa_{e} -1/2\right) }\\
     &\times\int_{-\infty }^{\infty }\frac{\left(1+x^{2}/\kappa_{e} \right) ^{-\kappa_{e}}}{x-g_{e}^{\pm }}dx,\ \  \Im \left(g _{e}^{\pm }\right) >0, \label{7}
     \end{aligned}
\end{equation}
of argument
\begin{equation}
g_{e}^{\pm }=\frac{\omega \pm \Omega _{e}}{k\theta _{\parallel, e}}.
\end{equation}
%
\begin{figure}[t]
\figurenum{1}
  \centering
   \includegraphics[width=7cm]{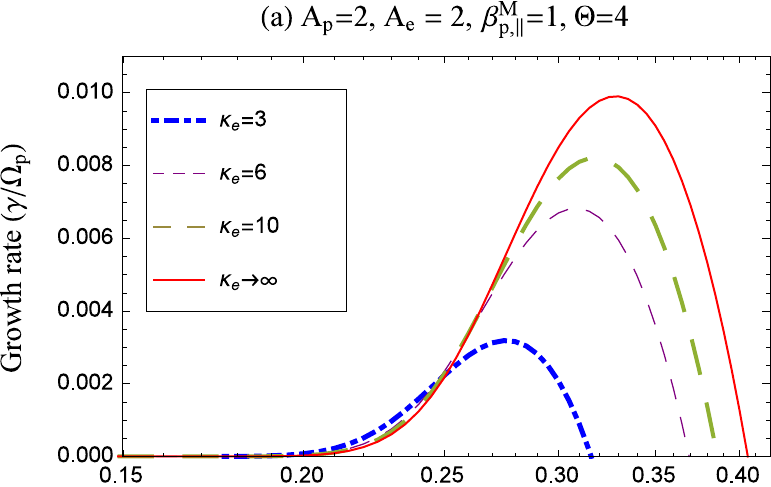}
   \includegraphics[width=7cm]{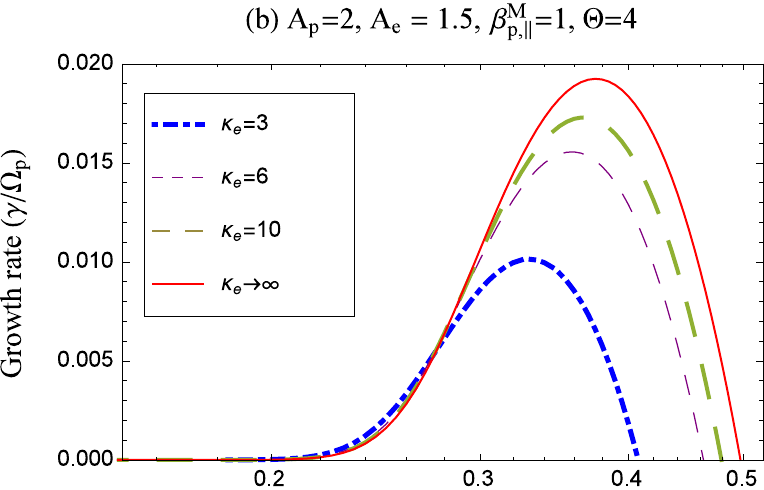}
   \includegraphics[width=7cm]{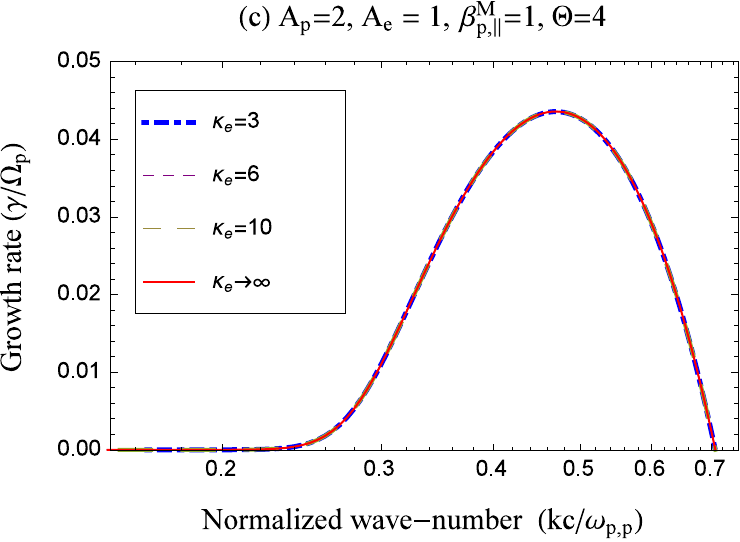}
\caption{Effects of the suprathermal electrons with ${\kappa}_{e}{=3,6,10,\infty}$ and $A_{e}=$~ 2 (top), 1.5~(middle), 1 (bottom) on the growth rates of EMIC instability for $A_{p}=2$, $\beta_{p,\parallel}^M=1$, $\Theta=4$.}\label{Fig1} 
\end{figure}
\begin{figure}[t]
\figurenum{2}
   \centering
   \includegraphics[width=7cm]{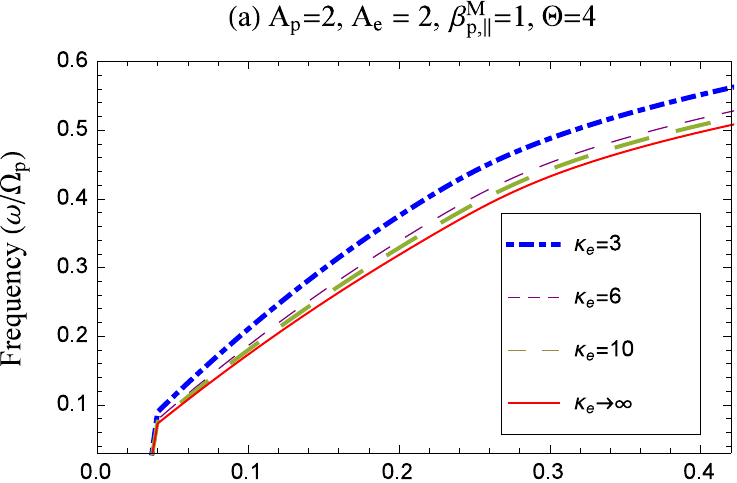}
   \includegraphics[width=7cm]{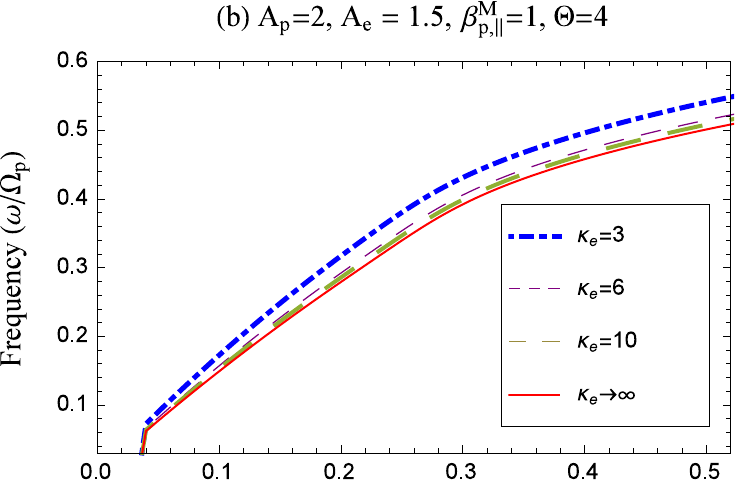}
   \includegraphics[width=7cm]{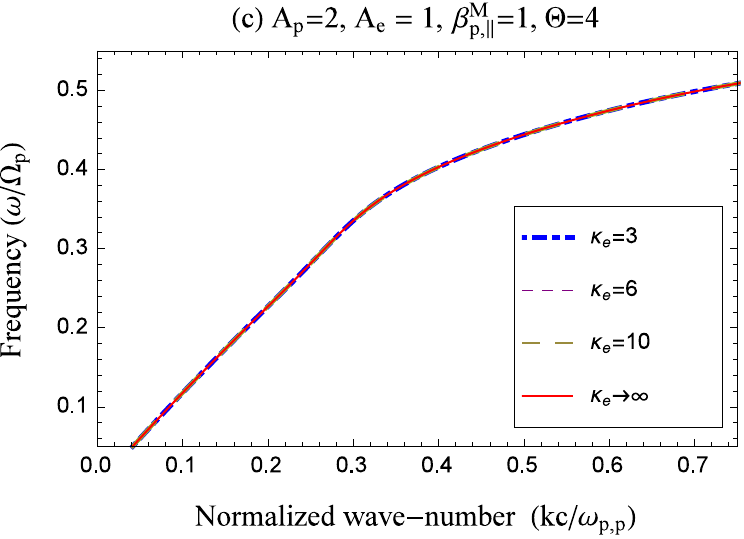} \label{Fig2} \caption{Effects of the suprathermal electrons with ${\kappa}_{e}{=3,6,10,\infty}$ and $A_{e}=$~ 2 (top), 1.5~(middle), 1 (bottom) on the wave-frequency of EMIC instability for $A_{p}=2$, $\beta _{p,\parallel}^M=1$, $\Theta=4$.}
\end{figure}
\begin{figure}[t]
\figurenum{3} 
 \centering
   \includegraphics[width=7cm]{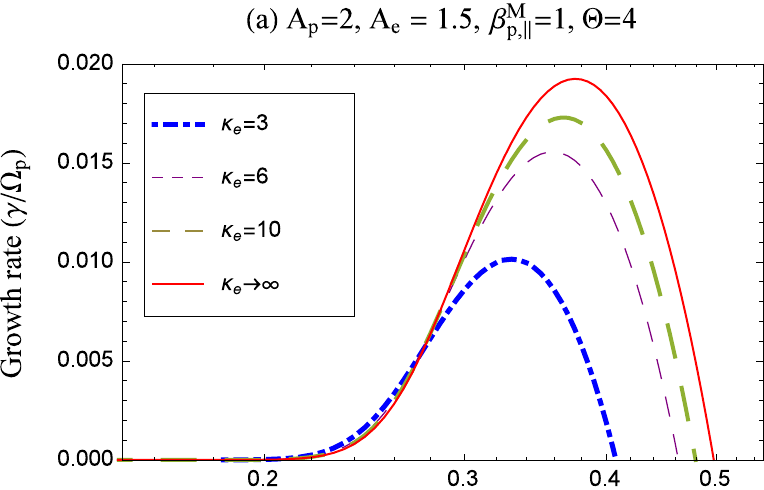}
   \includegraphics[width=7cm]{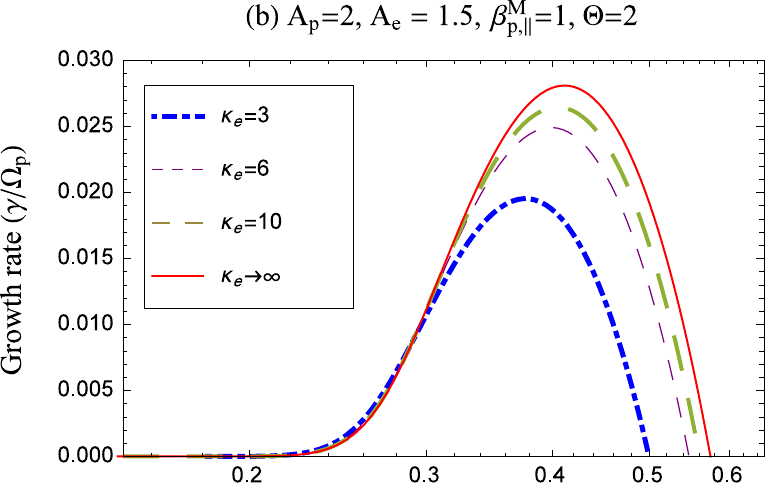}
   \includegraphics[width=7cm]{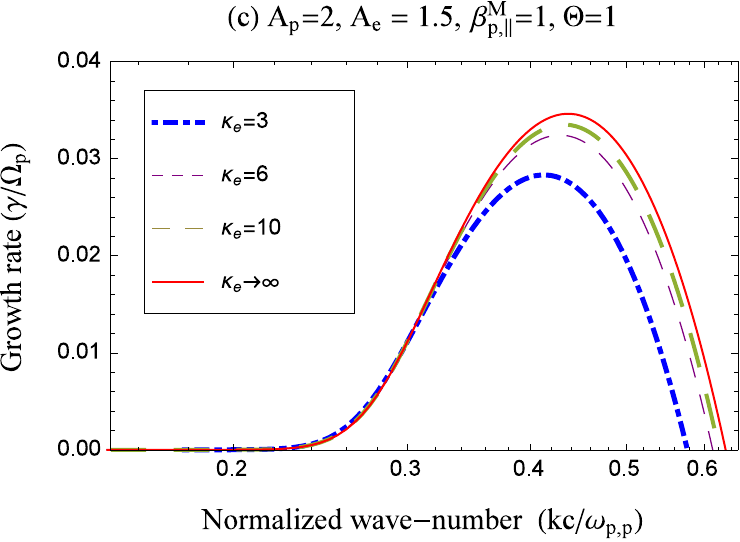} \label{Fig3} \caption{Effects of the suprathermal electrons with ${\kappa}_{e}{=3,6,10,\infty}$ and the temperature ratio $\Theta=$ 4~(top), 2~(middle), 1 (bottom) on the growth rates of EMIC instability for $A_{p}=2$, $A_{e}=1.5$, $\beta _{p,\parallel}^M=1$.}
\end{figure}
\begin{figure}[t]
\figurenum{4}  \centering
   \includegraphics[width=7cm]{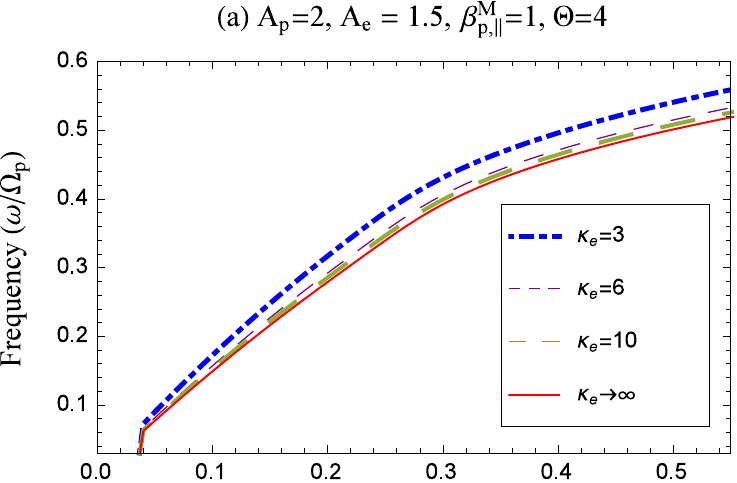}
   \includegraphics[width=7cm]{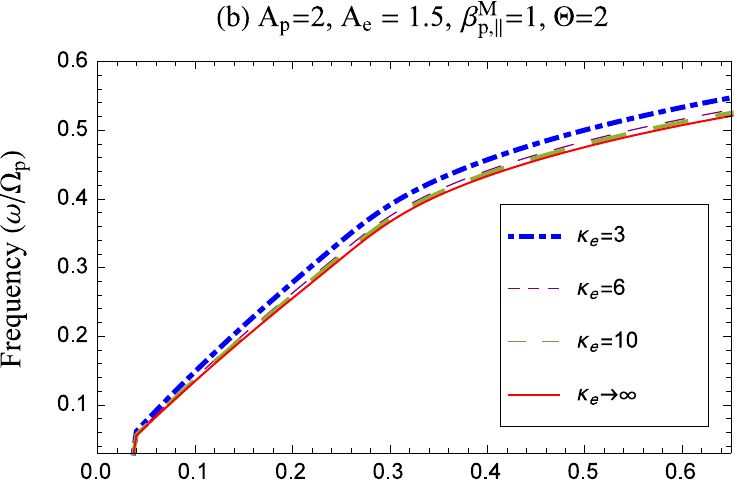}
   \includegraphics[width=7cm]{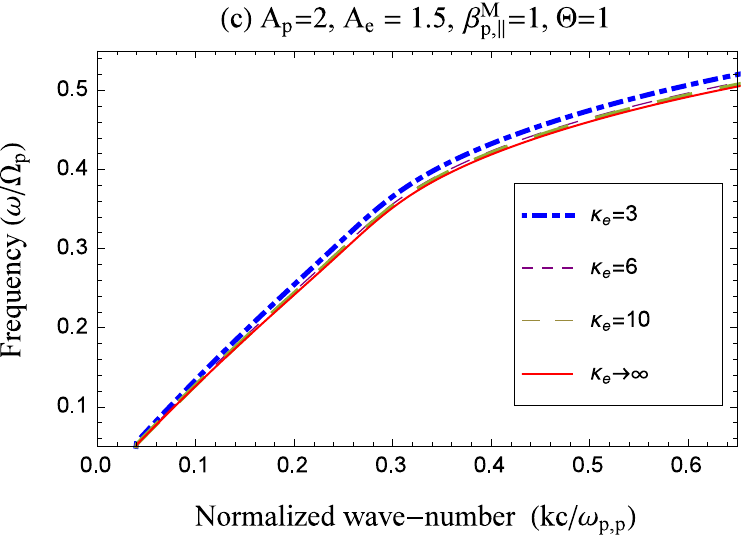} \label{Fig4} \caption{Effects of the suprathermal electrons with ${\kappa}_{e}{=3,6,10,\infty}$ and the temperature ratio $\Theta=$ 4~(top), 2~(middle), 1 (bottom) on the wave-frequency of EMIC instability for $A_{p}=2$, $A_{e}=1.5$, $\beta _{p,\parallel}^M=1$.}
\end{figure}
\begin{figure}[t]
\figurenum{5}  \centering
   \includegraphics[width=7cm]{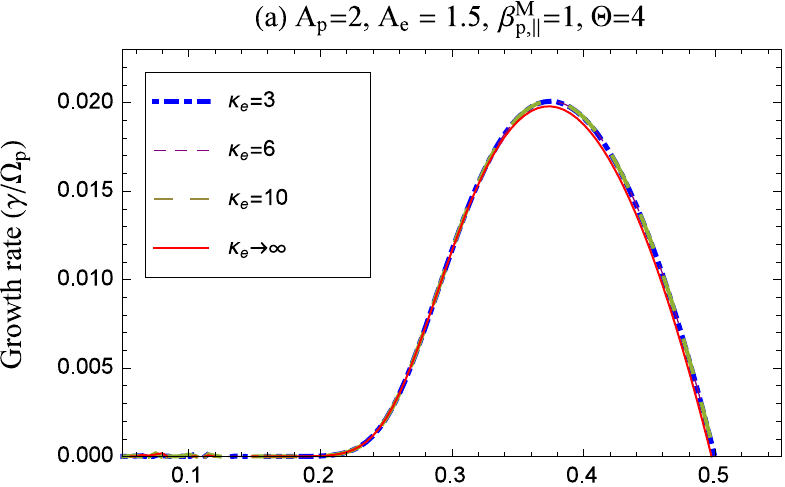}
   \includegraphics[width=7cm]{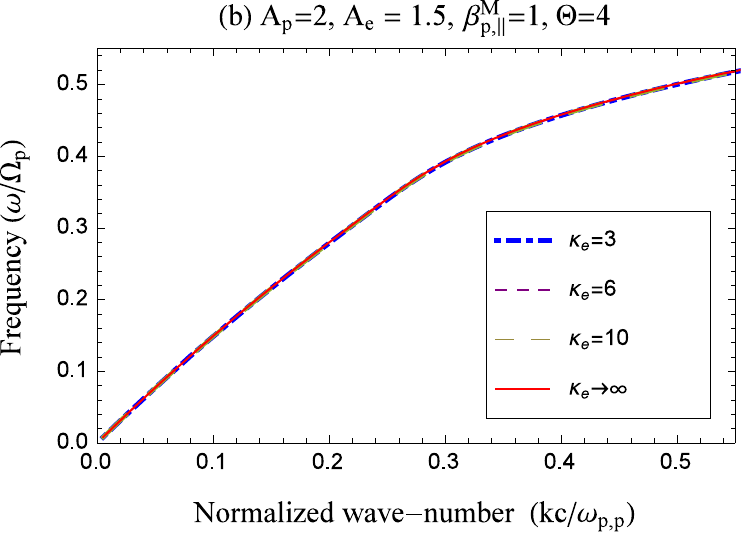}
    \label{Fig5} \caption{Effects of the suprathermal electrons with ${\kappa}_{e}{=3,6,10,\infty}$ on the growth rates (a) and wave-frequency (b) of EMIC instability for $A_{p}=2$, $A_{e}=1.5$, $\beta _{p,\parallel}^M=1$, $\Theta=4$.}
\end{figure}

For the EMIC modes, which have LH polarization and $\omega <~\Omega _{p}$,
the dispersion relation from Eq. (\ref{e9}) can be rewritten with
normalized quantities as follows
\begin{equation}
     \begin{aligned}
\mu \left(A_{e}-1\right)+\mu \frac{A_e \tilde{\omega}+\mu \left(A_e-1 \right)}{\tilde{k}\sqrt{\mu \Theta \beta_{p,\parallel}^M}}Z_\kappa\left( \frac{\tilde{\omega}+\mu }{\tilde{k}\sqrt{\mu \Theta \beta _{p,\parallel}^M}}\right)\\     
    +A_{p}-1-\tilde{k}^{2}+\frac{ A_{p}\left( \tilde{\omega}-1\right)+1 }{\tilde{k}\sqrt{\beta _{p,\parallel}^M}}Z\left( \frac{\tilde{\omega}-1}{\tilde{k}\sqrt{\beta_{p,\parallel}^M}}\right)=0
    \label{8}   
     \end{aligned}     
\end{equation}
where $\tilde{\omega}=\omega /\Omega _{p}$,
$\tilde{k}=~kc/\omega_{p,p}$, $\mu =~m_{p}/m_{e}$ is the
proton/electron mass ratio, $\Theta=~$~$T_{e,\parallel}^M/T_{p,\parallel }^M\ $ is
the electron/proton parallel temperature ratio in the Maxwellian
limit for both species, and  $\beta_{p, \parallel}^M = 8 \pi
n_{e} k_{B} T_{p,\parallel}^M / B_{0}^{2}$ is the parallel proton beta
parameter.  In the Maxwellian limit $\kappa \to \infty$ the dispersion relation (\ref{8}) reduces
exactly to Eq. (4) from \cite{Shaaban2015}.


\section{RESULTS}

The exact solutions of the dispersion relation (\ref{8}) are derived numerically,
providing accurate description for the unstable EMIC modes. We investigate the effects of the 
suprathermal electrons on the instability for two distinct cases, complementary to each other, 
i.e., either when the electrons exhibit an excess of perpendicular temperature, i.e., $A_{e}
> 1$, or the opposite case, when the electrons are more energetic
in direction parallel to the magnetic field, i.e., $A_{e} < 1$.

\subsection{\textit{Electrons with $A_{e}>1$}}

We first examine the case when the electrons show an excess of
transverse temperature, i.e., $T_{e,\perp} > T_{e,\parallel}$ (or
$A_e > 1$). In the absence of suprathermal populations,
\cite{Shaaban2015} have found that the electrons with $A_e
> 1$ inhibit the EMIC instability and this effect is stimulated
by the temperature ratio $\Theta$. Here we introduce a new
parameter, namely, the power-index $\kappa_e$ that quantifies the
presence and, implicitly, the effects of the suprathermal electrons. Figs.~\ref{Fig1} and
\ref{Fig2} display the wave-number dispersion for the growth rate,
and the wave-frequency, respectively, for the same set of parameters
$A_{p} =2$, $\beta_{p,\parallel}^{M}=1$, and $\Theta=4$, but
different kappa indices $\kappa_{e} = 3, 6, 10, \infty$ and different
electron anisotropies $A_{e}=1, 1.5, 2$. In Figs.~\ref{Fig3} and
\ref{Fig4} we keep constant the electron anisotropy $A_e = 1.5$ but
vary the electron/proton temperature ratio $\Theta=1, 2, 4$. All these
values are chosen according to the observations in the solar
wind \citep{stverak2008, Newbury1998}.

The inhibiting effect of the electron anisotropy on the EMIC instability is reconfirmed 
by the unstable solutions displayed in Fig.~\ref{Fig1}. In addition, the same figure shows 
that this effect may be significantly enhanced by the suprathermal electrons, namely, the growth-rate peaks
markedly decrease with decreasing the power-index $\kappa_e$. For isotropic electrons,
i.e., panels (c) in both Figs.~\ref{Fig1} and \ref{Fig2}, the suprathermal
populations do not have any influence on the EMIC instability. On the other hand, from a
comparison with Fig.~\ref{Fig3} we can observe that the EMIC instability is inhibited by increasing 
the temperature contrast $\Theta$ between electrons and protons, and again, this effect is 
stimulated by the suprathermal electrons (i.e., decreasing the power-index $\kappa_e$).
The wave-frequency in Figs.~\ref{Fig2} and \ref{Fig4} show an
opposite tendency, with values being slightly diminished by these effects.

For the sake of comparison, we have also studied the influence of the suprathermal
electrons on the EMIC instability in the alternative approach which assumes the 
temperature independent of kappa index. Fig.~\ref{Fig5} displays unstable solutions 
representative for this approach, and these solutions show a negligible influence of 
the suprathermal electrons. We have found the same insignificant influence on the  
instability thresholds, and so, there is no need to plot them here.

\begin{table}[t]
\centering \caption{Fitting parameters for the proton anisotropy in
Eq. (15)} \label{t1}
\begin{tabular}{c c c c }
\hline\hline
& & $\gamma_{\rm m}/\Omega = 10^{-3}$&  \\
 $\kappa_e$ & $A_{e}$& $a$  &$b$ \\
\hline
2        & 2 & 1.4043 & 0.1390 \\
         & 1.5 & 1.0852 & 0.1952 \\
         & 1 & 0.4504 & 0.4020\\
6        & 2 & 0.7624 & 0.2785 \\
         & 1.5 & 0.6295 & 0.3235 \\
         & 1 & 0.4504 & 0.4020\\
$\infty$ & 2 & 0.7028 & 0.2975 \\
         & 1.5 & 0.5920 & 0.3382 \\
         & 1 & 0.4504 & 0.4020\\
\hline
\end{tabular}
\end{table}


\begin{table}[t]
\centering \caption{Fitting parameters for the proton anisotropy in
Eq. (15)} \label{t2}
\begin{tabular}{c c c c }
\hline\hline
& & $\gamma_{\rm m}/\Omega = 10^{-2}$&  \\
 $\Theta$ & $\kappa_e$& $a$  &$b$ \\
\hline
4  & 2 & 1.7233 & 0.1742 \\
   & 6 & 1.0752 & 0.2840 \\
   & $\infty$ & 0.9903 & 0.3038\\
2  & 2 & 1.2980 & 0.2404\\
   & 6 & 0.8873 & 0.3304 \\
   & $\infty$ & 0.8352 & 0.3448\\
1  & 2 & 1.01818 & 0.2966 \\
   & 6 & 0.7784 & 0.3612 \\
   & $\infty$ & 0.7464 & 0.3711 \\
\hline
\end{tabular}
\end{table}

\begin{figure}[p]
\figurenum{6}
   \centering
   \includegraphics[width=5.8cm]{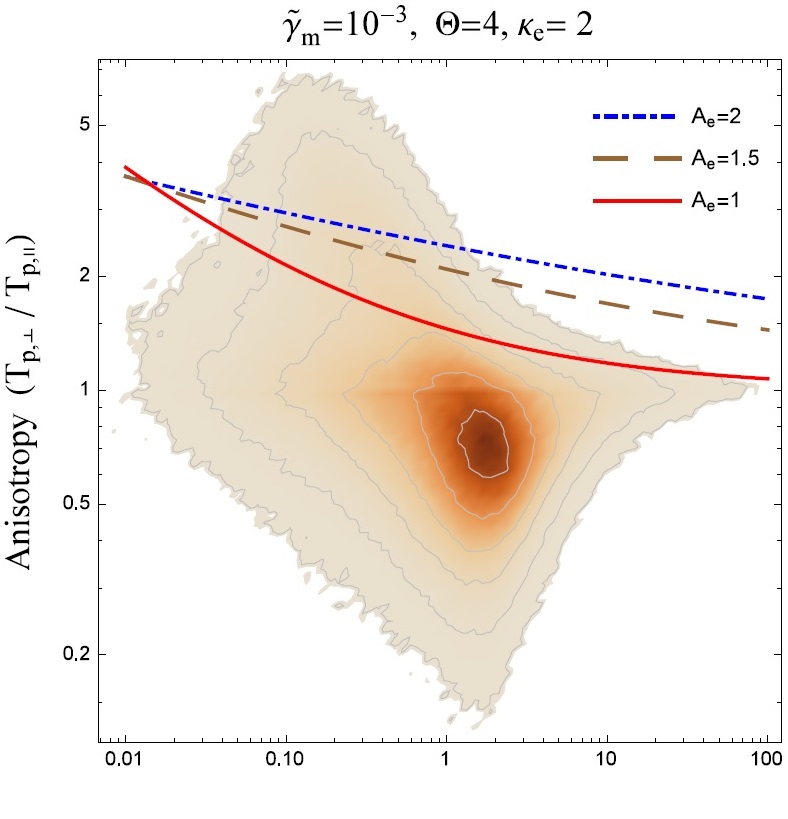}
   \includegraphics[width=5.8cm]{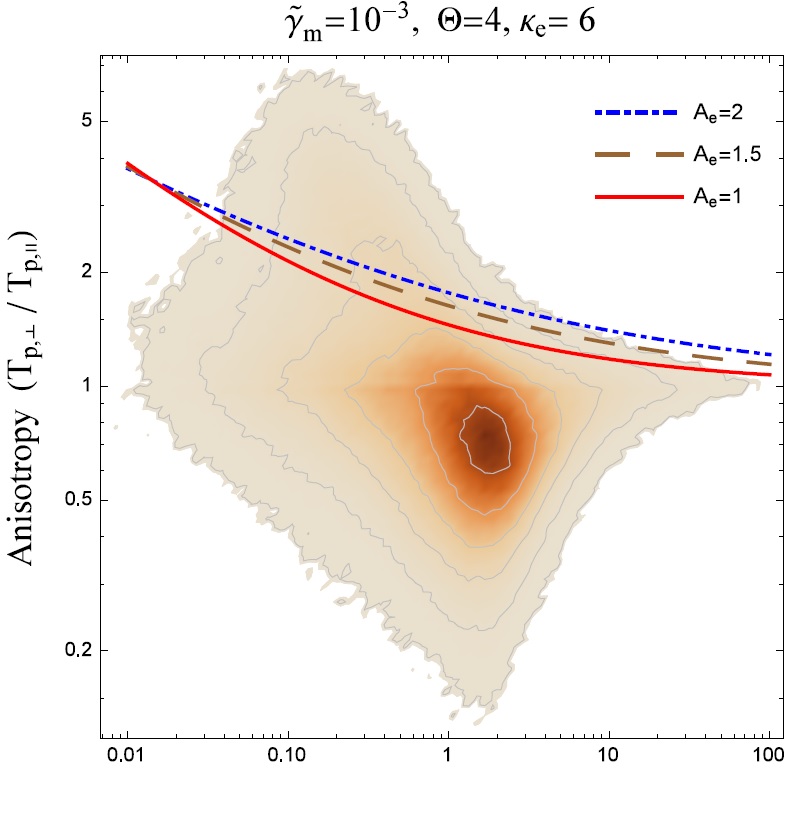}
   \includegraphics[width=5.8cm]{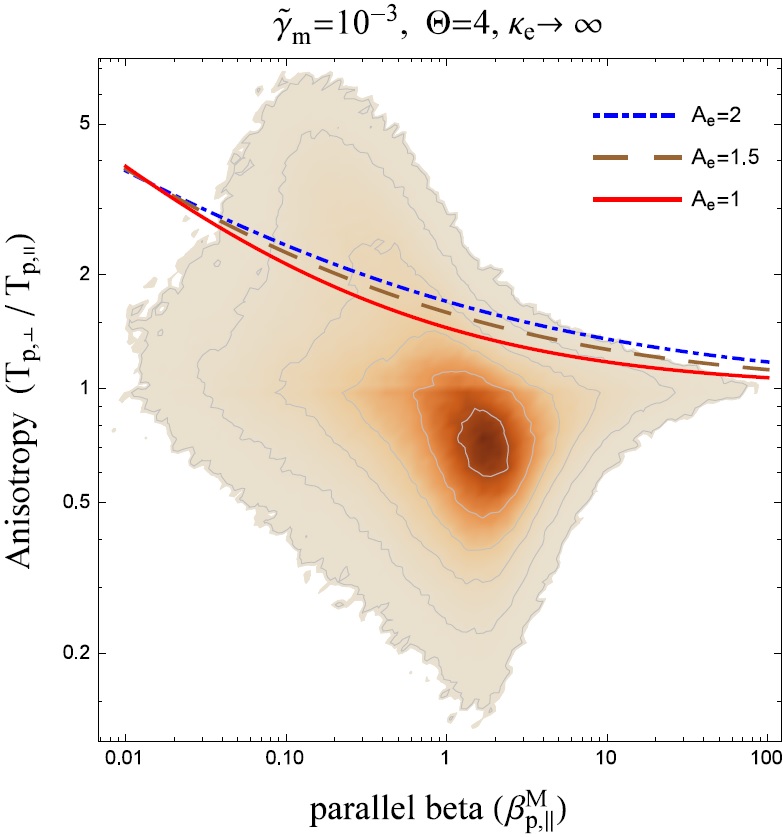}
   \includegraphics[width=5.8cm]{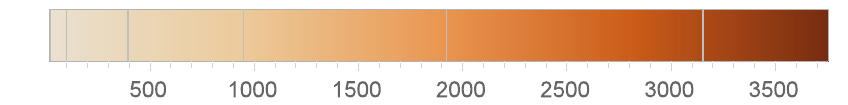}
\caption{Effects of the suprathermal electrons with ${A}_{e}{=2,1.5,1}$ and $\kappa_e=$ 2 (top), 6 (middle), $\infty$ (bottom) on the threshold conditions of EMIC instability with the maximum growth rate $\gamma_{m}/\Omega_{p}=~10^{-3}$ for $\Theta=4$.}
              \label{Fig6}%
    \end{figure}

\begin{figure}[p]
\figurenum{7}
   \centering
   \includegraphics[width=5.8cm]{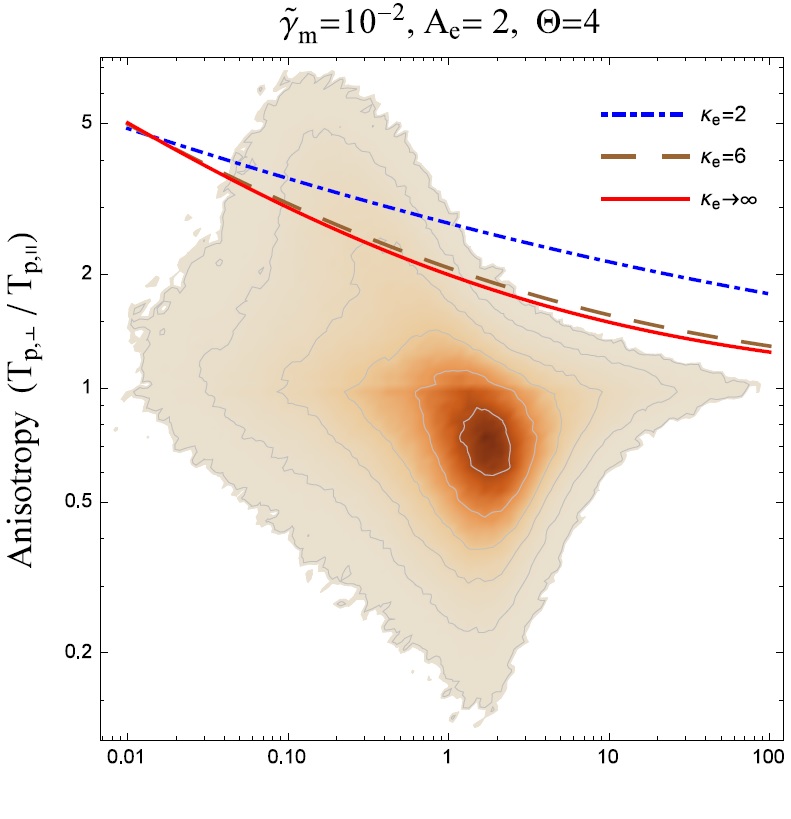}
   \includegraphics[width=5.8cm]{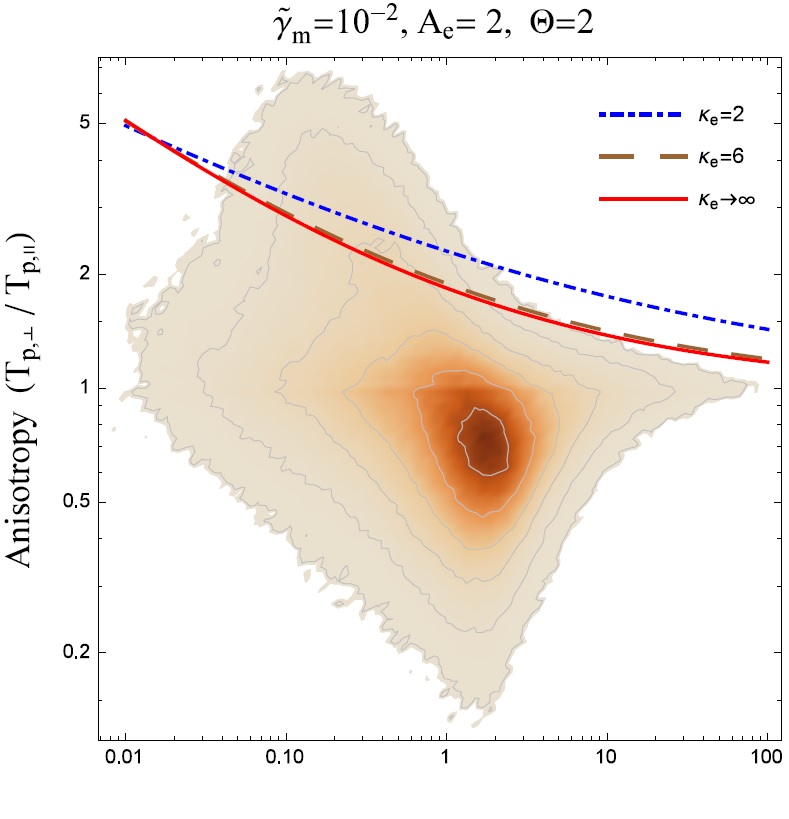}
   \includegraphics[width=5.8cm]{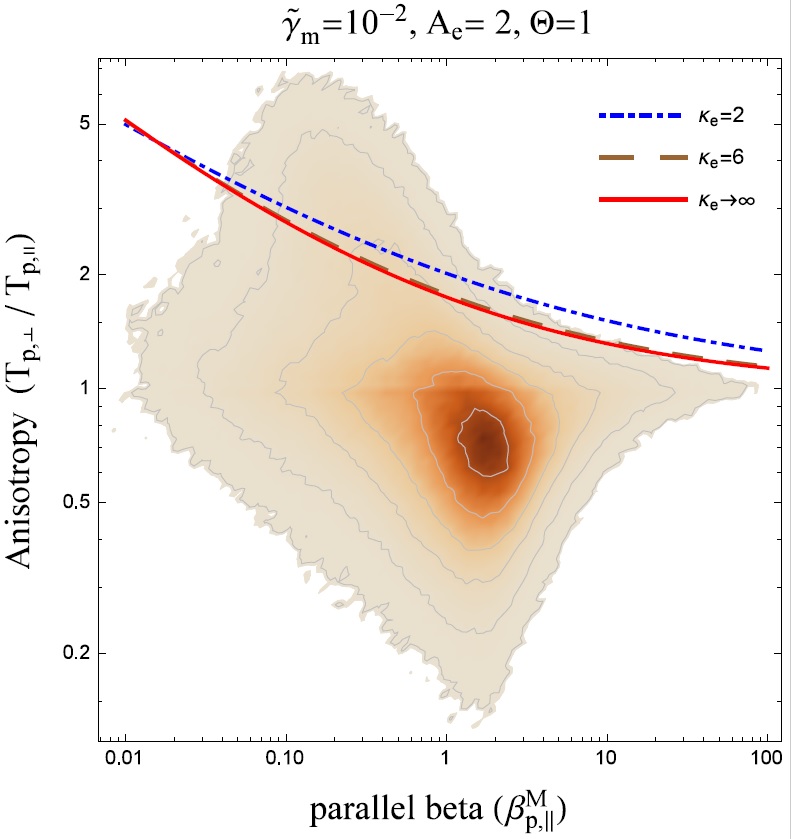}
   \includegraphics[width=5.8cm]{legend}
\caption{Effects of the suprathermal electrons with $\kappa_e=~2,6,\infty$ and the electron/proton temperature ratio $\Theta=$ 4~(top), 2 (middle), 1 (bottom) on the threshold conditions of EMIC instability with the maximum growth rate $\gamma_{m}/\Omega_{p}=~10^{-2}$ for $A_e=2$.}
              \label{Fig7}%
    \end{figure}

Finding a systematic inhibition of the EMIC instability under the effect of suprathermal
electrons, determined us to re-evaluate the anisotropy thresholds of this 
instability in the new conditions, and compare them with the limits of
the proton anisotropy observed in the solar wind. The anisotropy
thresholds for the EMIC instability are provided by the dispersion
relation for low levels of maximum growth-rates $\gamma_{\rm
m}/\Omega_p = 10^{-2}, 10^{-3}$, approaching marginal condition of
stability ($\gamma_{\rm m} = 0$). In Figs.~\ref{Fig6} and \ref{Fig7}
the anisotropy thresholds are calculated for an extended range of
the plasma beta parameter $0.01 < \beta_{p,\parallel}^M < 100$,
which includes conditions typical for the solar wind and terrestrial
magnetosphere. In Fig.~\ref{Fig6} we keep constant the temperature
ratio $\Theta = 4$, but vary the electron anisotropy $A_e = 2, 1.5, 1$
and the power-index $\kappa_e = 2, 6, \infty$. Contours of the maximum
growth rates $\gamma_{\rm m}/\Omega_{p} = 10^{-3}$ are fitted to an
inverse correlation law between the temperature anisotropy, $A_{p}$, and
the parallel plasma beta $\beta_{p,\parallel}^{M}$ \citep{Gary1994}
\begin{equation}
A_{p}=1+\dfrac{a}{{\beta_{p,\parallel}^M}^b}
\end{equation}
where the fitting parameters \textit{a} and \textit{b} are tabulated
in Table 1. The anisotropy thresholds plotted in Fig.~\ref{Fig6} are
increased by the electron anisotropy and this difference becomes more 
pronounced for higher values of the plasma beta parameter. These effects 
are found to be markedly enhanced by the suprathermal populations 
(i.e., decreasing the power-index
$\kappa_e$) confirming the influence shown already in
Figs.~\ref{Fig1} and \ref{Fig3} on the instability growth-rates. 


In Fig.~\ref{Fig7} we keep constant the electron anisotropy $A_e =
2$ and study the effect of the power-index $\kappa_e= 2, 6, \infty$
and the temperature ratio $\Theta = 4, 2, 1$ on the anisotropy
thresholds derived for $\gamma_{\rm m}/\Omega_{p}=10^{-2}$. Fitting
parameters \textit{a} and \textit{b} are tabulated in Table 2. The
effect of suprathermal electrons on the EMIC thresholds increases
with the temperature ratio $\Theta$, and this effect is still
apparent for isothermal species with $\Theta=1$.

\begin{figure}[t]
\figurenum{8}  \centering
   \includegraphics[width=7cm]{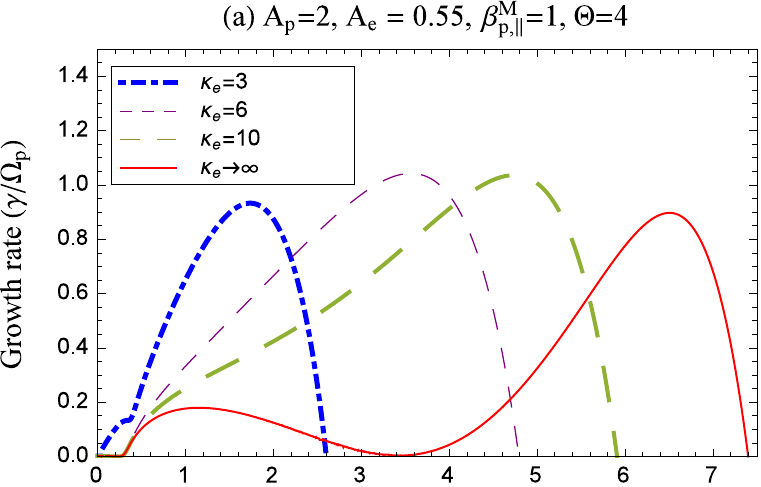}
   \includegraphics[width=7cm]{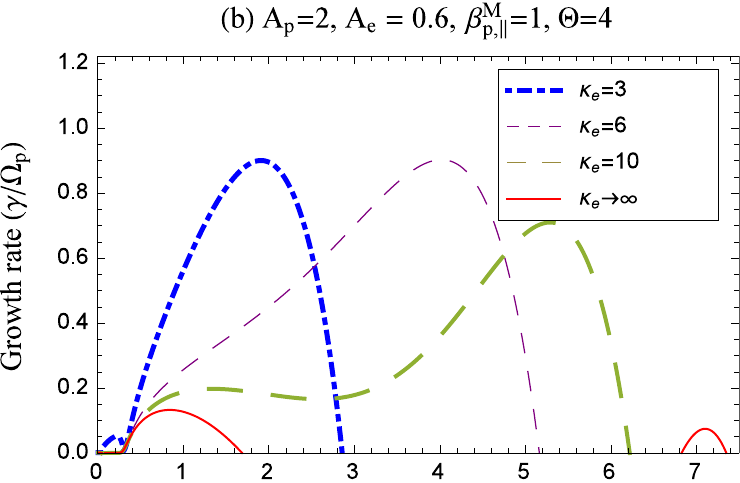}
   \includegraphics[width=7cm]{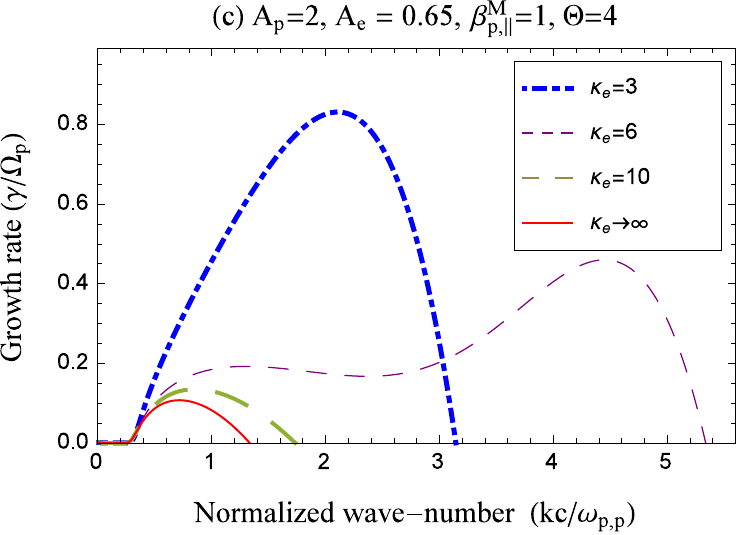} \label{Fig8} \caption{Effects of the suprathermal electrons with $\kappa_{e}{=3,6,10,\infty}$ and ${A}_{e}$~= 0.55 (top), 0.6 (middle), 0.65 (bottom) on the growth rates of EMIC instability for $A_{p}=2$, $\beta _{p,\parallel}^{M}=1$, $\Theta=4$.}
\end{figure}
\begin{figure}[t]
\figurenum{9}  \centering
   \includegraphics[width=7cm]{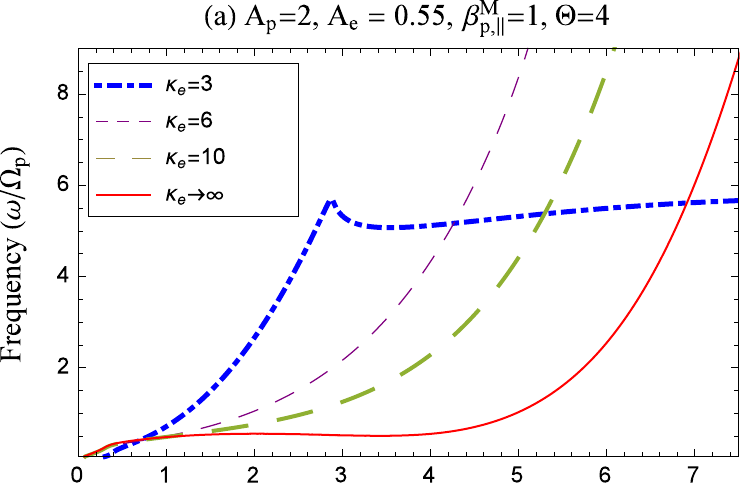}
   \includegraphics[width=7cm]{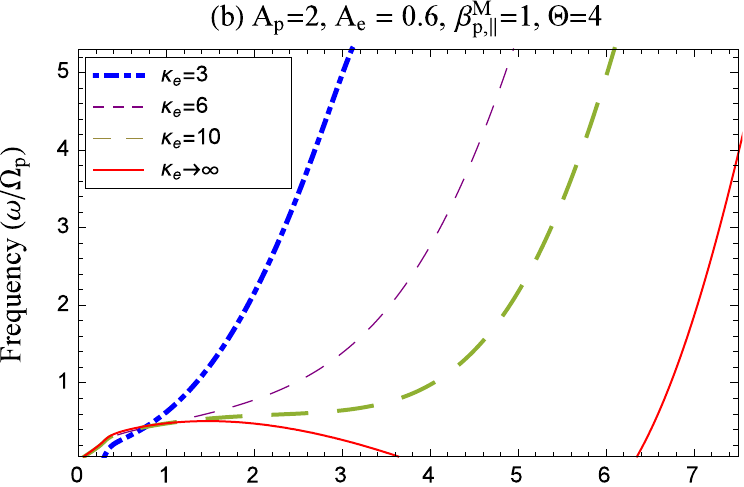}
   \includegraphics[width=7cm]{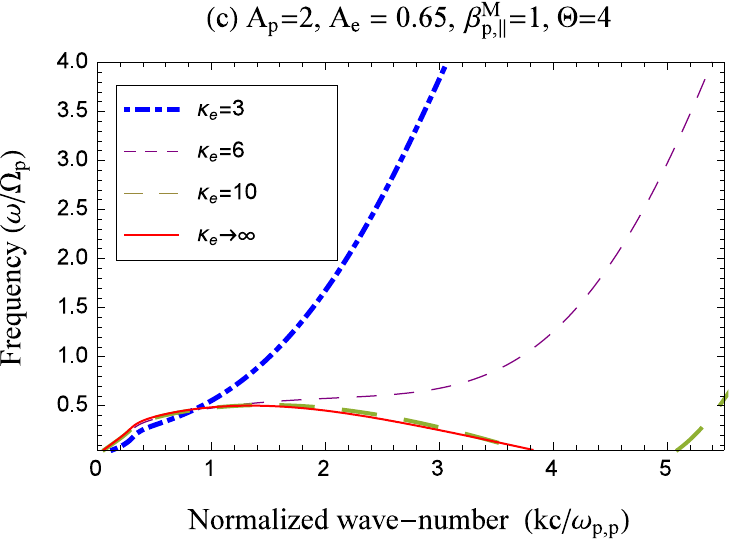}\label{Fig9} \caption{Effects of the suprathermal electrons with $\kappa_{e}{=3,6,10,\infty}$ and ${A}_{e}$~= 0.55 (top), 0.6 (middle), 0.65 (bottom) on the wave-frequency of EMIC instability for $A_{p}= 2$, $\beta _{p,\parallel}^{M}=1$, $\Theta=4$.}
\end{figure}

In both cases from Figs.~\ref{Fig6} and \ref{Fig7} the anisotropy
thresholds are compared with the limits of the proton anisotropy
from the 10-year measurements at 1~AU in the solar wind \citep{Bale2009, 
Shaaban2015}. We use the observational data from WIND/SWE and
MFI instruments for the proton velocity distribution and magnetic
field, respectively \citep{Lepping1995, Ogilvie1995}. The solar
wind proton anisotropy is displayed with a color logarithmic scale
representing the number of events (only for bins with more than 20
events).
According to our discussion in the introduction, the EMIC instability
is expected to constrain the proton anisotropy in the solar wind,
with instability thresholds aligned to (Gary et al. 2001) or
exceeding (Isenberg et al. 2013) the limits of the proton anisotropy
observed in the solar wind. In these new regimes investigated 
here, the electrons with anisotropic temperature and an important suprathermal 
component may have an important effect on the EMIC instability, 
inhibiting the growth-rates and increasing the thresholds, especially when
the instability is triggered by highly hot and anisotropic protons. 
However, the new EMIC thresholds obtained here 
do not show a good fit to the observations, but may rather exceed 
the limits of the proton anisotropy for large enough values of the 
plasma beta $\beta_{p,\parallel}^M$.

\subsection{\textit{Electrons with $A_{e}<1$}}

\begin{figure}[t]
\figurenum{10}  \centering
   \includegraphics[width=7cm]{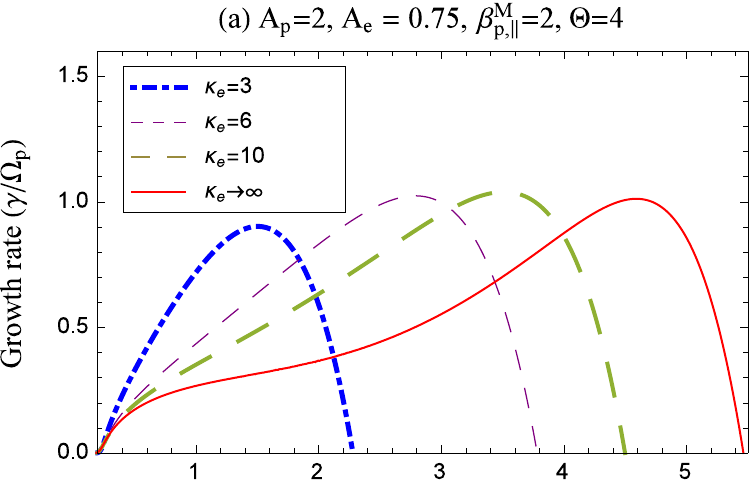}
   \includegraphics[width=7cm]{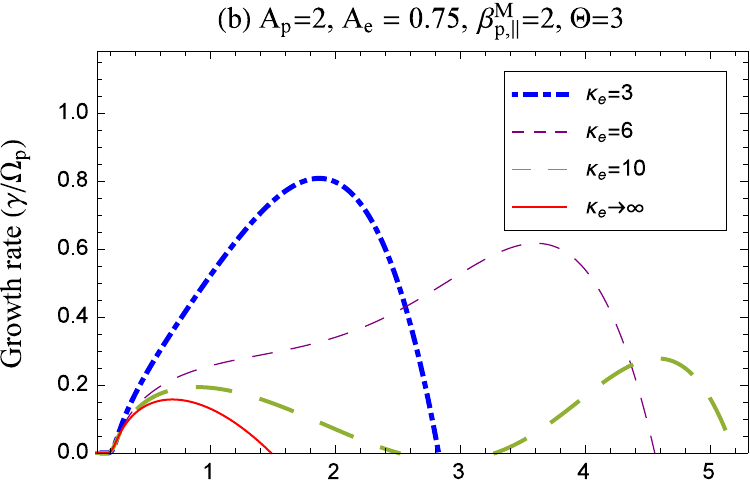}
   \includegraphics[width=7cm]{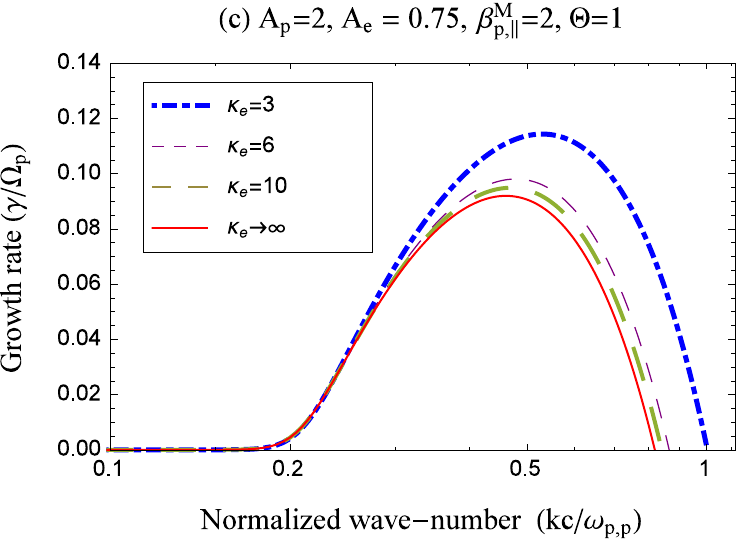} \label{Fig10} \caption{Effects of the suprathermal electrons with ${\kappa}_{e}{=3,6,10,\infty}$ and the temperature ratio $\Theta=$ 4~(top), 3 (middle), 1 (bottom) on the growth rates of EMIC instability for $A_{p}=2$, $A_e=0.75$, $\beta _{p,\parallel}^M=2$.}
\end{figure}
\begin{figure}[t]
\figurenum{11}  \centering
   \includegraphics[width=6.9cm]{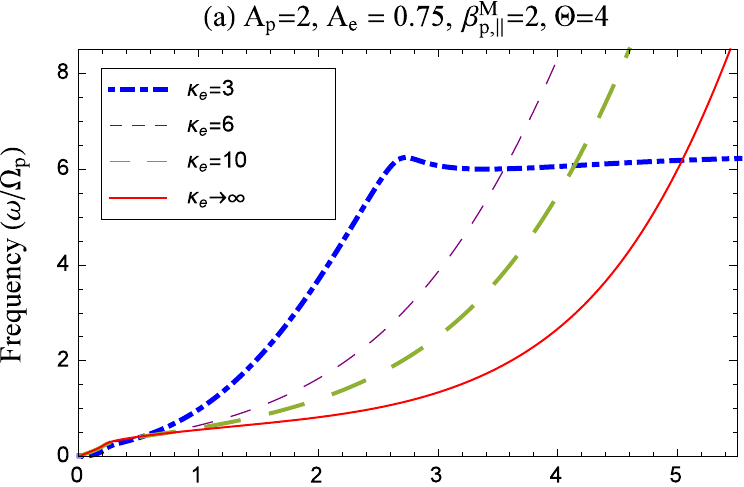}
   \includegraphics[width=6.9cm]{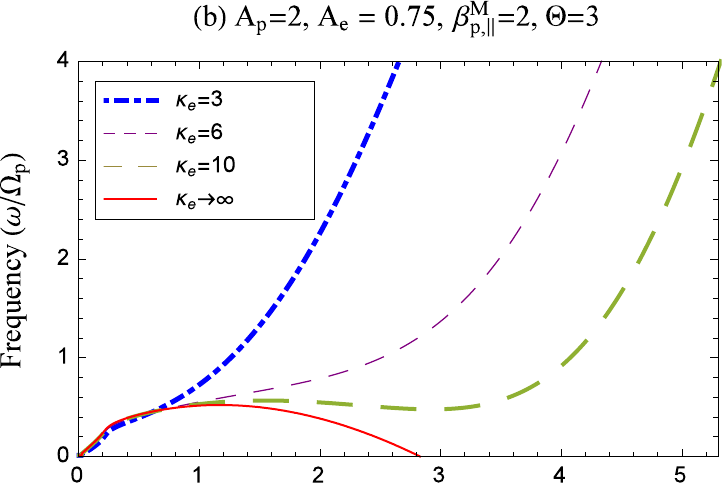}
   \includegraphics[width=6.9cm]{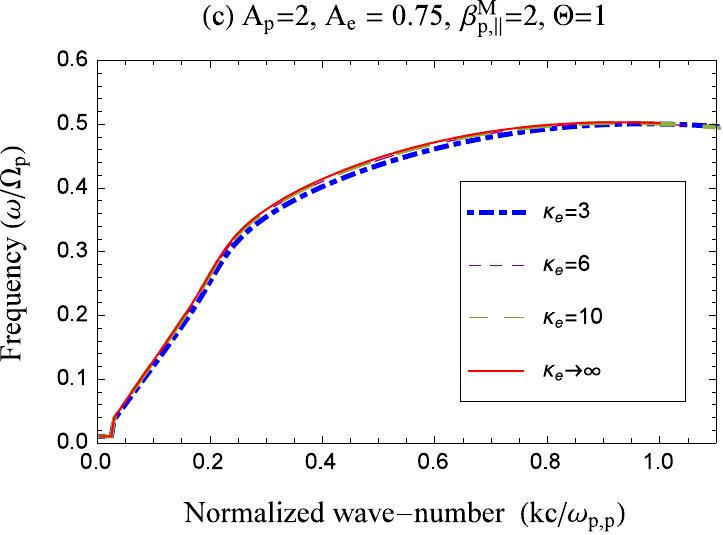}\label{Fig11} \caption{Effects of the suprathermal electrons with ${\kappa}_{e}{=3,6,10,\infty}$ and the temperature ratio $\Theta=$ 4~(top), 3 (middle), 1 (bottom) on the wave-frequency of EMIC instability for $A_{p}=2$, $A_e=0.75$, $\beta _{p,\parallel}^M=2$.}
\end{figure}
\begin{figure}[t]
\figurenum{12} \epsscale{0.9}  \centering
   \includegraphics[width=7cm]{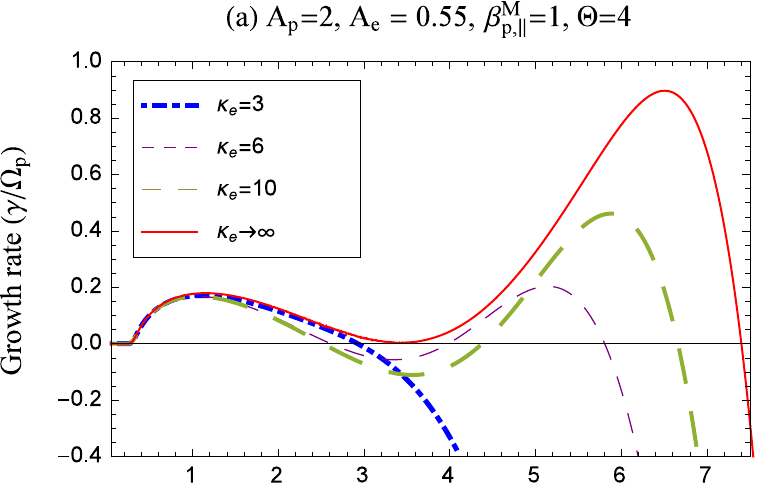}
   \includegraphics[width=7cm]{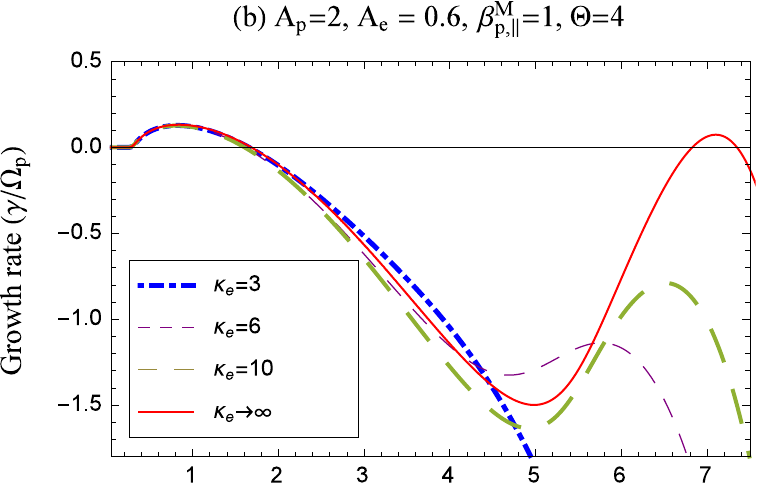}
   \includegraphics[width=7cm]{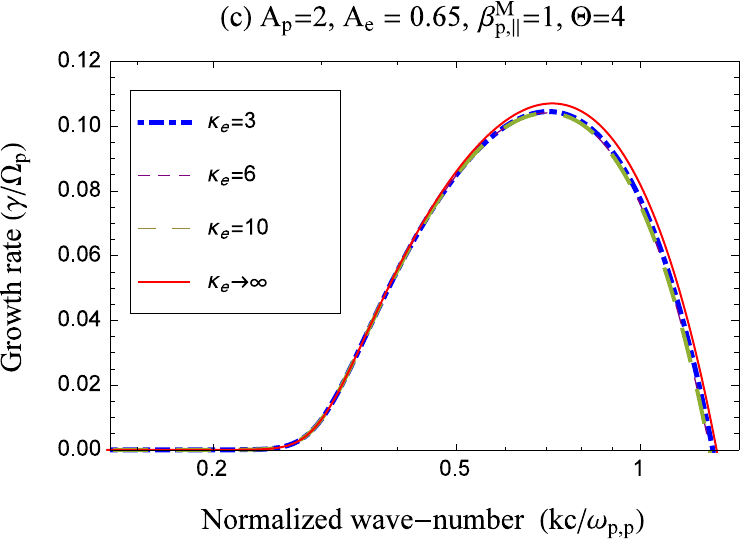} \label{Fig12} \caption{Effects of the suprathermal electrons with $\kappa_{e}{=3,6,10,\infty}$ and ${A}_{e}$~= 0.55 (top), 0.6 (middle), 0.65 (bottom) on the growth rates of EMIC instability for $A_{p}=2$, $\beta _{p,\parallel}^{M}=1$, $\Theta=4$.}
\end{figure}
\begin{figure}[t]
\figurenum{13}  \centering
   \includegraphics[width=6.8cm]{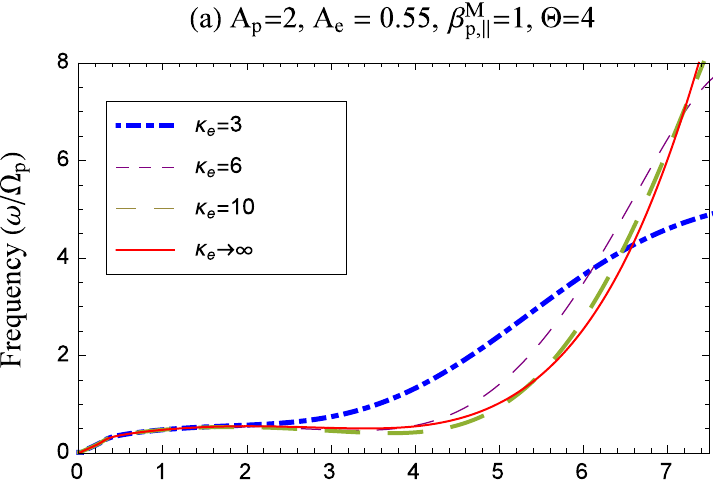}
   \includegraphics[width=6.8cm]{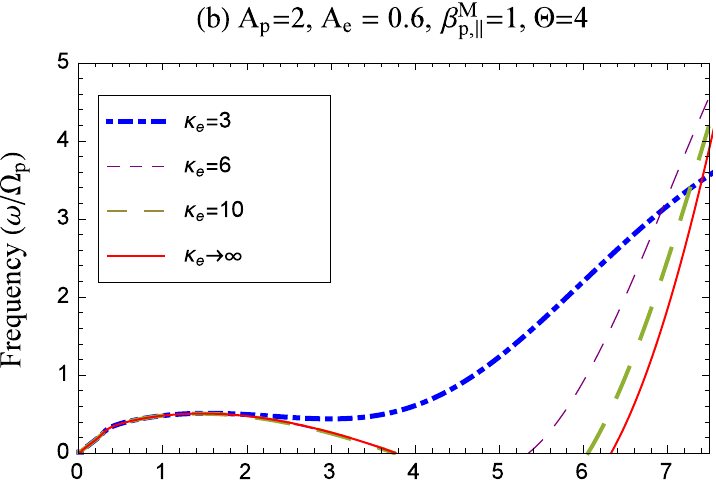}
   \includegraphics[width=6.8cm]{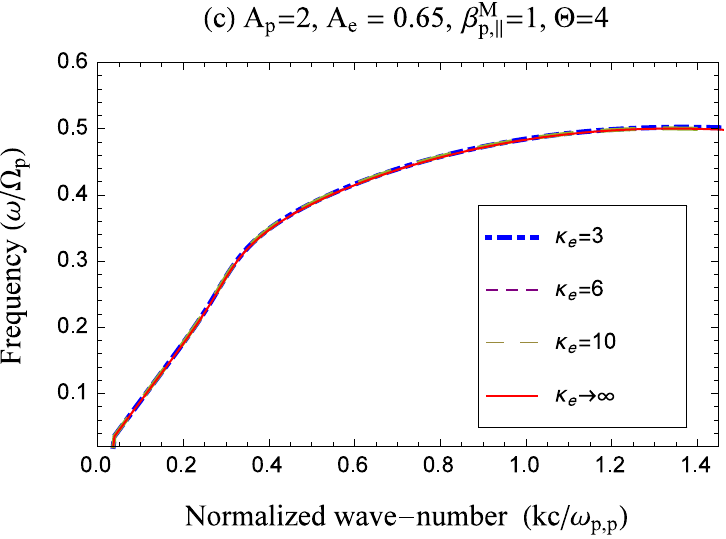} \label{Fig13} \caption{Effects of the suprathermal electrons with $\kappa_{e}{=3,6,10,\infty}$ and ${A}_{e}$~= 0.55 (top), 0.6 (middle), 0.65 (bottom) on the wave-frequency of EMIC instability for $A_{p}= 2$, $\beta _{p,\parallel}^{M}=1$, $\Theta=4$.}
\end{figure}
In the second part of this section we describe the influence of suprathermal electrons on the
EMIC instability in the opposite case, namely, when $T_{e,\parallel}
> T_{e,\perp}$. Figs.~\ref{Fig8} and \ref{Fig9} display the
wave-number dispersion of the growth rate and the wave-frequency,
respectively, for the same set of parameters $A_{p} =2$,
$\beta_{p,\parallel}^{M}=1$, and $\Theta=4$, but different kappa
indices $\kappa_{e}=~3, 6, 10, \infty$ and electron anisotropies
$A_{e}= 0.55, 0.6, 0.65$. In Figs.~\ref{Fig10} and \ref{Fig11} we
keep constant the electron anisotropy $A_e = 0.75$ but vary the
electron/proton temperature ratio $\Theta = 1, 3, 4$.

In Figs.~\ref{Fig8} and \ref{Fig10} the growth rates of the unstable
solutions may display two distinct peaks, one at low wave-numbers
corresponding to the EMIC mode, and a second peak arising at higher
wave-numbers due to the electron firehose (EFH) instability. Both
these two modes have the same LH circular polarization, which is
also confirmed by the same positive sign of their wave-frequencies in
Figs.~\ref{Fig9} and \ref{Fig11}. The existence and the dominance of
the second peak depend on both the electron anisotropy $A_e$ and the
power index $\kappa_e$. In the presence of suprathermal electrons
the EFH becomes dominant with a peak much higher than the EMIC, even
for small small deviations of the electrons from isotropy. For
instance, for the lowest values of the power-index $\kappa_e=3, 6$
in panels (a) and (b), the EMIC peak can only hardly be
distinguished. The unstable solutions obtained in the absence of 
suprathermal electrons ($\kappa_e \to \infty$) are plotted with solid lines, and their
dependence on the electron anisotropy and the electron/proton
temperature ratio is similar to that described by \cite{Shaaban2015}.

Another factor that stimulates the EFH instability is the
electron/proton temperature ratio $\Theta$ and for higher values
$\Theta = 3, 4$ the EMIC peak is completely hidden by the EFH peak,
see in Fig.~\ref{Fig10}, panels (a) and (b). However, the effect of the
suprathermal electrons on the EMIC instability (first peak) appears
more clear in panel (c), when the temperature contrast is minimized
($\Theta = 1$) and the deviation from isotropy is sufficiently small
($A_e = 0.75$). A detailed analysis of the EMIC thresholds by
comparison to the observations in the solar wind is difficult to
perform in this case, especially when the EMIC peak cannot be
distinguished from the EFH peak (see also the explanations in
\cite{Shaaban2015}). When the EMIC peak is distinguishable it is
found to be always stimulated by the suprathermal electrons with
$A_e < 1$, giving us possibility to conclude that the EMIC
thresholds are lowered and more deviated from the anisotropy limits
of solar wind protons.

Corresponding to these two instabilities, the wave-frequency
displayed in Figs.~\ref{Fig9} and \ref{Fig11} shows similar important
variations function of the electron anisotropy, the power index
$\kappa_e$ and the temperature contrast between electrons and protons.
For isothermal components, i.e., when $\Theta=1$, these variations
are negligibly small. These new results complement those obtained by
\cite{Lazar2011} and \cite{Michno2014} in studies of the firehose
instability cumulatively driven by the anisotropic protons and
electrons, when both species manifest an excess of parallel
temperature ($T_\parallel > T_\perp$). In that case the unstable
solutions show growth rates with two distinct peaks corresponding to
the proton firehose and electron firehose instabilities, and the
wave-frequency changes its sign according to the opposite
polarizations of these two modes.

We have studied the same plasma conditions invoked in Figs.~\ref{Fig8} and \ref{Fig9} in the alternative 
approach which assumes the effective temperature of the electrons independent of kappa index. 
For  the unstable solutions derived in this case, a number of representative cases are displayed in 
Figs.~\ref{Fig12} and \ref{Fig13}, showing the growth-rate and wave-frequency, respectively. 
In Figs.~\ref{Fig12} we find the EFH peaks (at larger wave numbers) significantly inhibited by the 
suprathermal  electrons, while the EMIC peaks (at lower wave-numbers) 
are not affected.  Moreover, with decreasing the electron anisotropy
the second peak is completely suppressed, and only the EMIC peak remains apparent (panel c). 
The wave-frequency may be significantly altered in the presence of suprathermal electrons 
but only at large wave-numbers corresponding to the EFH instability, see the plots in Figs.~\ref{Fig13}.


\section{DISCUSSIONS AND CONCLUSIONS}

The EMIC fluctuations are regularly observed in the solar wind, but their origin 
is still unclear. These are small scale fluctuations that can result from a decay of the
large scale fluctuation transported by the solar wind, or can be
generated locally by the EMIC instability driven by the temperature
anisotropy of protons (ions). The EMIC instability is fast enough to
constrain the proton anisotropy in a local scenario, but the
observations do not conform to the anisotropy thresholds predicted
by the early theories with simplified approaches. Recently,
\cite{Shaaban2015} have shown that the anisotropic electrons may
have important effects on the EMIC instability and this effect is
stimulated by the electron/proton temperature ratio. In this paper 
we have investigated for the first time the effects of the suprathermal 
electrons on this instability. The fluxes of suprathermal electrons are 
ubiquitous and highly anisotropic in the solar wind, and may, in general, be 
more intense than suprathermal protons.

Quantified by the power-index $\kappa_e$, the supra-thermal electrons 
are found to have an important influence on the EMIC instability, significantly 
enhancing the effects of the anisotropic electrons and the temperature ratio 
$\Theta$ previously described by \cite{Shaaban2015}. Thus, the electrons with $A_{e} > 1$ have an
inhibiting effect on the EMIC grow-rates \citep{Shaaban2015}, and
here we have shown that this effect is enhanced by the suprathermal
electrons: growth rates decrease with the decrease of the
power-index $\kappa_e$. Moreover, the effect of suprathermal
electrons is highly dependent on the electron anisotropy, vanishing
completely for isotropic electrons ($A_e=1$). On the other hand, the
inhibiting effects of the electron anisotropy and suprathermals
increase with the temperature ratio $\Theta$. All these effects are
confirmed by the EMIC thresholds derived in Figs.~\ref{Fig6} and
\ref{Fig7} for very low levels of the maximum growth rates
approaching the marginal stability. The EMIC thresholds obtained in
this case are increased by the electron anisotropy. This effect is
enhanced by the temperature contrast between electrons and protons
and by the suprathermal electrons, especially for higher values of
the proton plasma beta, i.e., $\beta_{p,\parallel} > 0.1$.
Comparison with the observations do not indicate a better alignment
of the EMIC thresholds to the limits of the proton anisotropy in the
solar wind. However, under the influence of suprathermal electrons, 
the EMIC thresholds show a tendency to exceed
the observational limits of the proton anisotropy, especially for
higher values of the plasma beta $\beta_{p,\parallel}^M > 1$, where
the potential of this instability to constrain the proton anisotropy
is significantly increased.

In the opposite case, the anisotropic electrons with $A_{e}<1$ stimulate  
the EMIC instability, and this effect is enhanced in the presence of suprathermal electrons 
(i.e., by decreasing the $\kappa$-index). However, a major enhancing effect 
is more apparent in this case for the EFH instability that gives rise to a second peak of the 
growth rates at higher wave-numbers. Thus, the EFH peak may in general exceed the EMIC peak,
which becomes indistinguishable for certain conditions. However, the EMIC growth rates are 
always enhanced by the suprathermal electrons with $A_e < 1$, leading to the conclusion 
that the EMIC thresholds are diminished in this case, and are even less relevant for the anisotropy limits of solar
wind protons. We must add that all these effects are revealed only by a Kappa approach 
which assumes the effective temperature of electrons increasing with the increase of suprathermal
electrons. Otherwise, for a Kappa model with the effective temperature independent of $\kappa_e$, 
the EMIC instability is not affected by these populations.

Our results in the present paper complement those of a recent series of investigations on 
the proton anisotropy instabilities, namely, the parallel firehose \citep{Michno2014}, and the 
EMIC instability \citep{Shaaban2015}. These studies are based on advanced and less-idealized kinetic 
approaches, which enable us to decode the interplay of thermal (core) protons with thermal and suprathermal 
electrons, and implicitly their destabilizing effects on different plasma modes. Thus, the
parallel firehose instability is found to be markedly affected by the presence of anisotropic 
electrons, such that the instability thresholds predicted under some circumstances may describe the observations without considering the oblique firehose mode \citep{Michno2014}. The same
anisotropic electrons can also change the EMIC thresholds but whithout a satisfactory
reshaping that could explain the observations \citep{Shaaban2015}.
In conclusion, we now can add that suprathermal electrons have an important influence on the 
EMIC instability, and this influence is highly conditioned by two principal factors: 
the temperature anisotropy of electrons which may stimulate or inhibit the EMIC fluctuations, and
the electron/proton temperature ratio which enhances the effects of the anisotropic electrons.
Moreover, the EMIC instability thresholds undergo major changes in the presence of anisotropic 
electrons with $A_{e}<1$ and an important suprathermal component (i.e., for lower values of 
$\kappa_e$), but the new thresholds do not show better alignment to the limits of the proton 
anisotropy $A_p >1$ measured at 1 AU in the solar wind. Apparently negative, these results 
are however important, as they provide us with enlightening answers, namely, that neither 
the mutual electron-proton effects analyzed by \cite{Shaaban2015}, nor the presence of 
suprathermal electrons studied in the presence paper can influence the EMIC thresholds to 
explain the limit of proton anisotropy in the solar wind.

\begin{acknowledgements}
The authors acknowledge the use of WIND SWE (Ogilvie et al. 1995)
ion data, and WIND MFI (Lepping et al. 1995) magnetic field data
from the SPDF CDAWeb service: http://cdaweb.gsfc.nasa.gov/. The
authors acknowledge support from the Katholieke Universiteit Leuven.
These results were obtained in the framework of the projects
GOA/2015-014 (KU Leuven), G0A2316N (FWO-Vlaanderen), and C 90347
(ESA Prodex 9). The research leading to these results has also
received funding from IAP P7/08 CHARM (Belspo), and the European Commission's Seventh Framework Programme FP7-PEOPLE- 2010-IRSES-269299 project-SOLSPANET(www.solspanet.eu). This project has received funding 
from the European Union's Seventh Framework Programme for research, technological development and demonstration under grant agreement SHOCK 284515. S.M. Shaaban would like to thank the Egyptian Ministry of Higher Education for supporting his research activities and and would like to acknowledge the discussions and suggestions of Prof. S. A. Elwakil.
\end{acknowledgements}

\end{document}